# Towards radio astronomical imaging using an arbitrary basis

Matthias Petschow

Netherlands Institute for Radio Astronomy (ASTRON)

petschow@astron.nl

**Abstract**

The new generation of radio telescopes, such as the Square Kilometer Array (SKA), requires dramatic advances in computer hardware and software, in order to process the large amounts of produced data efficiently. In this document, we explore a new approach to wide-field imaging. By *generalizing* the image reconstruction, which is performed by an inverse Fourier transform, to *arbitrary transformations*, we gain enormous new possibilities. In particular, we outline an approach that might allow to obtain a sky image of size $P \times Q$ in (optimal) $\mathcal{O}(PQ)$ time. This could be a step in the direction of *real-time*, wide-field sky imaging for future telescopes.

## 1 Image synthesis via Fourier transformation

Traditionally, a radio astronomical image or sky intensity distribution $I^{(\nu)}(\ell, m)$, depending on angular coordinates $\ell, m \in [-1, 1]$, is derived from a (non-uniformly) sampled visibility function $V^{(\nu)}(u, v, w)$. For a specific frequency $\nu$ (or a small band around $\nu$), and making a set of common assumptions, the visibility $V^{(\nu)}(u, v, w)$ is connected to $I^{(\nu)}(\ell, m)$ by:

$$V^{(\nu)}(u,v,w) = \int_{-\infty}^{\infty} \int_{-\infty}^{\infty} I_{mod}^{(\nu)}(\ell, m) \, A^{(\nu)}(\ell, m) \, e^{-2\pi i(u\ell + vm + wn)} d\ell \, dm \,, \tag{1.1}$$

where $I_{mod}^{(\nu)}(\ell, m) = I^{(\nu)}(\ell, m)/\sqrt{1 - \ell^2 - m^2}$ denotes a modified intensity, $n = \sqrt{1 - \ell^2 - m^2} - 1$, and $A^{(\nu)}(\ell, m)$ is the *normalized antenna reception pattern* or *primary beam* [5, 6].

Each sample $V^{(\nu)}(u_{ij}(t), v_{ij}(t), w_{ij}(t))$ in the $(u, v, w)$-space is derived by measuring the cross correlation of antenna pair signals $(i, j)$, i.e. $\langle E^{(\nu)}(\vec{r}_i, t), E^{(\nu)*}(\vec{r}_j, t) \rangle$, where $\langle \cdot \rangle$ means time-averaging, $E^{(\nu)}(\vec{r}_i, t)$ is the electric field at position $\vec{r}_i$, and $*$ denotes conjugate transposition.[1] Evidently, for each measurement, we obtain two samples, as

$$V^{(\nu)}(u_{ij}(t), v_{ij}(t), w_{ij}(t)) = V^{(\nu)*}(-u_{ij}(t), -v_{ij}(t), -w_{ij}(t)).$$

Consequently, given $Y$ antennas, $2 \cdot \binom{Y}{2}$ samples are taken at any moment in time. The terms $(u_{ij}(t), v_{ij}(t), w_{ij}(t))$, the so called *baselines*, corresponds to $\vec{r}_i - \vec{r}_j$ measured in wavelength and are expressed in the $(u, v, w)$-coordinate system [5, 6]. The time-dependence arrives from tracking a source and thereby rotating the $(u, v, w)$-coordinate system.

Several strategies justify the reduction of (1.1) to a two-dimensional Fourier transformation: only a small region in the sky is observed and consequently $n \approx 0$, all antennas are almost in a plane and consequently $w \approx 0$, or the term $e^{-2\pi i w n}$ is treated as a convolution of $V^{(\nu)}(u, v, w = 0)$ [2, 6].

---

[1] (1) In this document, the word antenna stands for a single antenna or a beamformed station of antennas. (2) For simplicity, we consider the electric fields as scalar fields. An extension vector fields is possible.

If the beams $A^{(\nu)}(\ell, m)$ are different for each antenna pair, the convolution theorem is also used to correct for the $A$-terms, before a Fourier transform is performed [1].

The so called *dirty image* is obtained by a two-dimensional Fourier transform of the (non-uniformly) sampled (and corrected) visibility function. Further processing is needed to get a *clean image*, which is an approximation to the true sky intensity distribution $I^{(\nu)}(\ell, m)$: as sampling implies that the dirty image is a convolution of the true sky intensity with the *dirty beam* or *point spread function* (PSF) of the instrument, the process to obtain a clean image is also called *deconvolution*.

## 2 Image synthesis via an arbitrary transformation

### 2.1 The idea

For each antenna pair (baseline), after discretizing space $\vec{\ell} = (\ell_1, \ell_2, \ldots, \ell_P)$, $\vec{m} = (m_1, m_2, \ldots, m_Q)$ (for now, equidistant[2]) and time with $\vec{t}_{ij} = (t_{ij,1}, t_{ij,2}, \ldots, t_{ij,K_{ij}})$ (possibly baseline-dependent) in (1.1), we approximate the measured visibility at time $t_{ij,k}$ by

$$V^{(\nu)}(u_{ij}(t_{ij,k}), v_{ij}(t_{ij,k}), w_{ij}(t_{ij,k})) =$$
$$\sum_{p=1}^{N} \sum_{q=1}^{M} \Delta\ell \, \Delta m \, I_{mod}^{(\nu)}(\ell_p, m_q) \, A_{ij,k}^{(\nu)}(\ell_p, m_q) \, e^{-2\pi i (u_{ij}(t_{ij,k})\ell_p + v_{ij}(t_{ij,k})m_q + w_{ij}(t_{ij,k})n_{pq})} \,. \quad (2.1)$$

In (2.1), we dropped any assumption on the antenna pair beams $A^{(\nu)}$; now, they can be different for each antenna pair and time-dependent.[3]

The key insight to express the intensity function in a convenient basis $\{\Psi_\Omega\}$, with dual basis $\{\Psi_\Omega^D\}$, is to consider linear combinations of visibilities:

$$c^{(\nu)}(\hat{\Psi}_\Omega) = \sum_{i=1}^{Y} \sum_{\substack{j=1 \\ j \neq i}}^{Y} \sum_{k=1}^{K_{ij}} \sigma_{i,j,k} V^{(\nu)}(u_{ij}(t_{ij,k}), v_{ij}(t_{ij,k}), w_{ij}(t_{ij,k})) =$$
$$\sum_{p=1}^{P} \sum_{q=1}^{Q} \Delta\ell \, \Delta m \, I_{mod}^{(\nu)}(\ell_p, m_q) \underbrace{\sum_{i=1}^{Y} \sum_{\substack{j=1 \\ j \neq i}}^{Y} \sum_{k=1}^{K_{ij}} \sigma_{i,j,k} A_{ij,k}^{(\nu)}(\ell_p, m_q) \, e^{-2\pi i (u_{ij}(t_{ij,k})\ell_p + v_{ij}(t_{ij,k})m_q + w_{ij}(t_{ij,k})n_{pq})}}_{\text{new generating vector } \hat{\Psi}_\Omega^D(\vec{\ell}, \vec{m})} \,.$$
$$(2.2)$$

This leads to a new *generalized visibilities*, $\{c^{(\nu)}(\hat{\Psi}_\Omega)\}$, which are linear combinations of the regular visibilities (see Fig. 2.2); the generalized visibilities correspond to coefficients in the new generating set $\{\hat{\Psi}_\Omega\}$, where $\Omega$ is a parameter set (such as a discretized $(u, v)$-space or a set of indices). We get

---

[2]We assume for simplicity that the samples fill the entire sky. In a more general scenario, they only cover the region of interest. The discretization scheme warrants a more detailed discussion.

[3]We assume that the time-dependent effects are known a priori. These could be beam shape modulations or re-pointing of antennas.

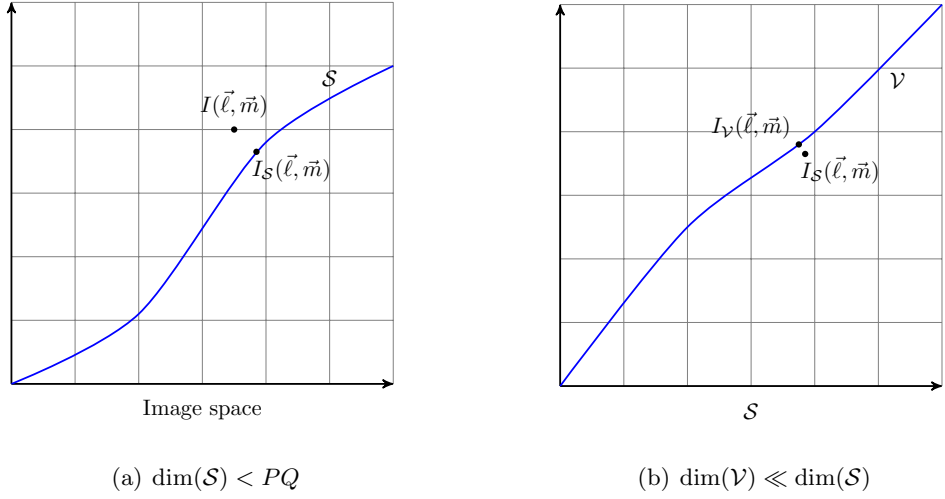

(a) $\dim(\mathcal{S}) < PQ$  (b) $\dim(\mathcal{V}) \ll \dim(\mathcal{S})$

Figure 2.1: If $\dim(\mathcal{S}) < PQ$, we can recover the projection onto approximation space $\mathcal{S}$, $I_{\mathcal{S}}(\vec{\ell}, \vec{m})$. If we have domain-specific knowledge about coefficients that are close to zero, we obtain a good approximation of the image by restricting ourselves to subspace $\mathcal{V} \subset \mathcal{S}$ with $\dim(\mathcal{V}) \ll \dim(\mathcal{S})$.

the following relation:[4]

$$\begin{aligned}C^{(\nu)}(\Psi_\Omega) &= \sum_{p=1}^{P}\sum_{q=1}^{Q} \Delta\ell\,\Delta m\, I_{mod}^{(\nu)}(\ell_p, m_q)\hat{\Psi}_\Omega^D(\ell_p, m_q) \\ &\approx \sum_{p=1}^{P}\sum_{q=1}^{Q} \Delta\ell\,\Delta m\, I_{mod}^{(\nu)}(\ell_p, m_q)\Psi_\Omega^D(\ell_p, m_q)\,.\end{aligned} \qquad (2.3)$$

Eq. (2.3) defines the transformation between coefficients $C^{(\nu)}(\Psi_\Omega) = \{c^{(\nu)}(\hat{\Psi}_\Omega)\}$ and the modified intensity $I_{mod}^{(\nu)}$:

$$C^{(\nu)}(\Psi_\Omega) = \mathcal{T}\{I_{mod}^{(\nu)}\}\,. \qquad (2.4)$$

In order to invert transformation $\mathcal{T}$ efficiently, say in $\mathcal{O}(PQ)$ time, we select a *suitable* basis $\{\Psi_\Omega\}$. Often, the basis is unitary, or real and orthogonal. To obtain meaningful images, "enough" dual basis vectors $\Psi_\Omega^D$ must be *well-approximated*, i.e.

$$\|\Psi_\Omega^D - \hat{\Psi}_\Omega^D\| < \bar{\delta}_\Omega \qquad \text{or} \qquad \frac{\|\Psi_\Omega^D - \hat{\Psi}_\Omega^D\|}{\|\Psi_\Omega^D\|} < \delta_\Omega \qquad (2.5)$$

for some norm $\|\cdot\|$ and ($\Omega$-dependent) thresholds $\bar{\delta}_\Omega$ or $\delta_\Omega$. As the $\{\Psi_\Omega\}$ basis spans the $PQ$-dimensional image space, we obtain a projection onto reduced approximation space $\mathcal{S}$, if only $\dim(\mathcal{S}) < PQ$ dual basis vectors satisfy (2.5); such a situation is visualized in Fig. 1(a). The coefficients for not well-represented basis vectors are set to zero.

---

[4]We do not discuss the approximation error in detail. However, we provide a way to bound the approximation error in Appendix A.

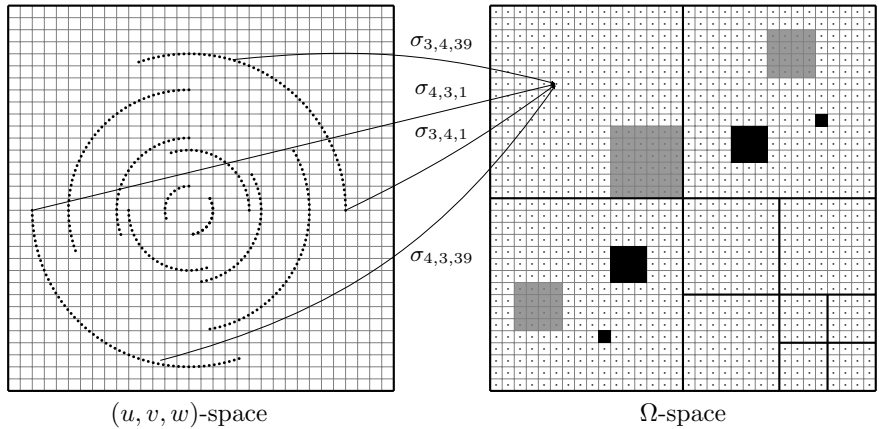

Figure 2.2: The new basis is designed such that a fast transformation is possible (so to speak "on a uniform grid"). Each coefficient is a linear combination of visibilities, given as non-uniform samples in the $(u, v, w)$-space. The black boxes indicate basis vectors that are not well approximated and whose coefficients cannot be obtained reliably. The gray boxes are coefficients that are known to be zeros due to application-knowledge. For an optimal basis, there would be very few black, but many gray boxes, i.e. $\dim(\mathcal{S}) \approx PQ$ and $\dim(\mathcal{V}) \ll \dim(\mathcal{S})$.

If we have chosen a basis, which is *well-represented* and is *suitable* for astronomical images, we are hopefully able to exploit further domain-specific knowledge to anticipate coefficients that are zero or close to zero (see gray boxes in Fig. 2.2). In such a case, the *computation is greatly reduced* and a good low-rank approximation $I_\mathcal{V}(\vec{\ell}, \vec{m})$ in $\mathcal{V} \subset \mathcal{S}$, with $\dim(\mathcal{V}) \ll \dim(\mathcal{S})$, of the sky brightness achieved. Such a scenario is visualized in Fig. 1(b).

## 2.2 Advantages and disadvantages of the new approach

In this section, we anticipate several advantages and disadvantages of the approach. Among the advantages are:

- The above *generalizes* the approach of using a non-uniform Fourier basis, which is still contained by selecting $\Psi_\Omega^D = \Psi^D(u_{ij}(t_{ij,k}), v_{ij}(t_{ij,k})) = e^{-2\pi i(u_{ij}(t_{ij,k})\ell + v_{ij}(t_{ij,k})m)}$, i.e. non-uniform points in the continuous $(u,v)$-plane. In this case, only one weight factor is equal to one, while the others are zero, and it is apparent from (2.2) that we either have to assume $n_{pq} = 0$ or $w_{ij}(t_{ij,k}) = 0$ (or deal with it using $W$-projection [2]). Furthermore, we have to assume $A_{ij,k}^{(\nu)}(\vec{\ell}) = 1$ or correct for the $A$-terms using $A$-projection [1].
- By designing a suitable basis so to speak on "a uniform grid", an *image is obtained efficiently by a simple inverse transformation*. There is *no need for gridding* the non-uniform samples in Fourier space to a equidistant grid in order to perform an FFT, and there is *no need to correct for the $W$-terms and $A$-terms* by convolutions. In particular, discrete wavelets bases are promising candidates. In this case, the inverse fast wavelet transformation (FWT) is performed in $\mathcal{O}(PQ)$ time. Furthermore, the wavelet basis can be *tailored to the demands of radio astronomical images and the observing instrument*.

However, before the approach becomes practical, several obstacles have to be addressed:

- The determination of weight factors $\{\sigma_{i,j,k}\}$ might be computationally intractable. Only in special cases, we might be able to determine the weights efficiently. In any case, we believe that the computation of the weights should be performed in a "offline" stage, instead of "online" during an observation.
- Given certain baselines, which of the desired bases can be well approximated, i.e., have $\dim(\mathcal{S}) \approx PQ$? Can we use beamforming to obtain the antenna beams that are suitable?
- Once we reconstructed an image using $\mathcal{T}^{-1}$, all the later processing stages need to be reexamined. In particular, for a Fourier basis, we know that the dirty image is the sky intensity convolved with the PSF of the instrument. Similarly, in the new approach, the weights define a PSF that must be approximated and used for later deconvolution. This computational step needs still to be investigated within the new approach.

## 2.3 How to approximate a basis

A challenging question is the determination of weights $\{\sigma_{i,j,k}\}$ such that (2.5) is satisfied for the basis vectors $\Psi_\Omega$ that are most relevant to an image. Defining $F^{(\nu)}_{ij,k}(\vec{\ell}, \vec{m})$ as the complex exponential $e^{-2\pi i (u_{ij}(t_{ij,k})\ell_p + v_{ij}(t_{ij,k})m_q + w_{ij}(t_{ij,k})n_{pq})}$, for each dual basis vector $\Psi^D_\Omega$, we need to determine

$$\arg\min_{\{\sigma_{i,j,k}\}} \|\Psi^D_\Omega(\vec{\ell}, \vec{m}) - \sum_{i,j,k} \sigma_{i,j,k} \underbrace{A^{(\nu)}_{ij,k}(\vec{\ell}, \vec{m}) \circ F^{(\nu)}_{ij,k}(\vec{\ell}, \vec{m})}_{\text{"measurement vectors"}}\| \tag{2.6}$$

for some norm $\|\cdot\|$, with element-wise Hadamard product $\circ$, and $\Psi^D_\Omega, A^{(\nu)}_{ij,k}, F^{(\nu)}_{ij,k}$ of size $P \times Q$. *For both efficiency and accuracy, the sum should generally be restricted to a (small) subset of possible values.* For later reference, we denote the terms $A^{(\nu)}_{ij,k} \circ F^{(\nu)}_{ij,k}$ as *measurement vectors*; therefore, we search for linear combinations of measurement vectors that represent the dual basis vectors well.

For instance, by choosing $\|\cdot\|_F$, (2.6) becomes a linear least squares problem (LLP). However, other requirements such as sparsity of the solution might be important, or even analytic solutions possible for a certain choice of $\{\Psi_\Omega\}$. The general approach for obtaining the weight factors is outlined in Algorithm 1. The inputs are a desired basis $\{\Psi_\Omega\}$, a set of baselines via $\{F^{(\nu)}_{ij,k}\}$, corresponding antenna beams $\{A^{(\nu)}_{ij,k}\}$, and tolerances $\{\delta_\Omega\}$.

---

**Algorithm 1** FINDWEIGHTS($\{\Psi_\Omega\}, \{F^{(\nu)}_{ij,k}\}, \{A^{(\nu)}_{ij,k}\}, \{\delta_\Omega\}$)

  **for** $\gamma \in \Gamma \subseteq \Omega$ **do**
    For dual basis vector $\Psi^D_\gamma$, solve (2.6) to obtain a set of weights $s_\gamma = \{\sigma_{i,j,k}\}$.
    **if** $\{s_\gamma, \Psi^D_\gamma\}$ satisfy (2.5) for $\delta_\gamma$ **then**
      Store the weights $s_\gamma$ for determination of coefficient $c^{(\nu)}(\hat{\Psi}_\gamma)$.
    **else**
      Mark $\Psi_\gamma$ as unfeasible by setting $c^{(\nu)}(\hat{\Psi}_\gamma) = 0$.
    **end if**
  **end for**

---

## 2.4 Compute costs

Ideally the entire imaging procedure requires only $\mathcal{O}(PQ)$ [at most, $\mathcal{O}(PQ \cdot \log_2 PQ)$] operations. This includes the creation of a dirty image, as well as the corrections for system noise and atmospheric effects. In this document, we only look a the first step. As there exist inverse transformations $\mathcal{T}^{-1}$ that only cost $\mathcal{O}(PQ)$ [or $\mathcal{O}(PQ \cdot \log_2 PQ)$] operations, we must be able to limit the computation of the coefficients to $\mathcal{O}(PQ)$ [or $\mathcal{O}(PQ \cdot \log_2 PQ)$] operations as well. *Assuming the weight factors are known a priori*, each of the generalized visibilities can only be a linear combination of $\mathcal{O}(1)$ [or $\mathcal{O}(\log_2 PQ)$] of visibilities. We give motivation below, why we believe that, in certain situations, this might be possible.[5] Once we have created a first image, the transformation between $\Omega$-space and image domain is performed in $\mathcal{O}(PQ)$ [or $\mathcal{O}(PQ \cdot \log_2 PQ)$] time. Any algorithm for corrections for system noise and atmospheric effects that performs this transformation a limited number of times, with $\mathcal{O}(PQ)$ [or $\mathcal{O}(PQ \cdot \log_2 PQ)$] work in-between, would achieve the desired result.

## 3 The simplified one-dimensional case

### 3.1 The new setting

From here on, we concentrate on the easier one-dimensional case, as we believe that the simplest case should be understood thoroughly first, before making any (possibly incorrect) conclusions that the approach is not feasible in practice or too computationally expensive.

In the new setting, the brightness distribution $I(\ell)$ only depends on one angular coordinate $\ell$ and, after discretization, baselines are given by $\{(u_{ij}(t_{ij,k}), w_{ij}(t_{ij,k}))\}$.[6] To further simplify the problem, we assume that the phase reference is at zenith, and the antennas are in a plane such that $w_{ij}(t_{ij,k}) = 0$ for all baselines and all time. The baseline for antenna $i$ and $j$ is now described by only its $u$-values during time, $\{u_{ij}(t_{ij,k})\}$. If the antenna responses $A_{ij,k}(\vec{\ell})$ are equal for all antenna pairs or time-independent, it does not matter if the samples are taken by one baseline at different times or many baselines at one moment in time. The sampling in $u$-space is simply given by a set of values $\{u_b\} = \{u_{-B}, \ldots, u_{-2}, u_{-1}, u_1, u_2, \ldots, u_B\}$, with $u_b = -u_{-b}$, and where $b$ is considered to be an index of the baseline or time. The corresponding antenna responses are denoted $A_b(\vec{\ell})$.

With all these simplifications, (2.1) becomes

$$V(u_b) = \sum_{p=1}^{P} \Delta\ell\, I(\ell_p)\, A_b(\ell_p)\, e^{-2\pi i u_b \ell_p}. \qquad (3.1)$$

For constant $A_b(\vec{\ell}) = A(\vec{\ell})$, (3.1) describes the connection between visibilities $V(\{u_b\})$ and *observed intensity* $I(\vec{\ell}) \circ A(\vec{\ell})$ as a non-uniform discrete Fourier transformation (NDFT). Forming linear combinations of the visibilities results in

$$c(\hat{\psi}_\Omega) = \sum_{\substack{b=-B \\ b \neq 0}}^{B} \sigma_b V(u_b) = \sum_{p=1}^{P} \Delta\ell\, I(\ell_p) \underbrace{\sum_{\substack{b=-B \\ b \neq 0}}^{B} \sigma_b A_b(\ell_p)\, e^{-2\pi i u_b \ell_p}}_{\text{new generating vector } \hat{\psi}_\Omega^*(\vec{\ell})} \approx \sum_{p=1}^{P} \Delta\ell\, I(\ell_p)\, \psi_\Omega^*(\ell_p), \qquad (3.2)$$

---

[5]We simply assume that it is possible. It has still to be shown that this assumption is valid.
[6]Here and in the following, we drop all the dependence on frequency $\nu$. Furthermore, the intensity $I(\ell)$ corresponds to the modified intensity, i.e. the true intensity is $I(\ell)\sqrt{1-\ell^2}$.

where we assume that $\{\psi_\Omega\}$ is a unitary basis. (This assumption is made throughout the rest of the document.) Eq. (3.2) corresponds to (2.2) and defines transformation $\mathcal{T}$ between generalized visibilities and intensity distribution as in (2.4): $C(\psi_\Omega) = \mathcal{T}(I)$.

## 3.2 Determining the weight factors

To determine weight factors $\{\sigma_b\}$, we need to minimize the norm of the residual $r_\Omega = \psi_\Omega - \hat{\psi}_\Omega$, i.e. solve

$$\arg\min_{\{\sigma_b\}} \|\psi_\Omega^*(\vec{\ell}) - \sum_b \sigma_b \underbrace{A_b(\vec{\ell}) \circ e^{-2\pi i u_b \vec{\ell}}}_{\text{"measurement vectors"}}\| =: \arg\min_{\vec{\sigma}} \|\psi_\Omega^* - s^* M_\Omega\|, \quad (3.3)$$

where $M_\Omega \in \mathbb{C}^{L \times P}$, $2 \leq L \leq 2B$, with a subset of measurement vectors as rows, and $s \in \mathbb{C}^L$ containing the weights. Depending on the ratio of the number of samples in the sky, $P$, and the number of samples in $u$-space, $L$, the system can be over-determined or under-determined. As $B$ grows as fast as $\binom{Y}{2}$, for $Y$ antennas, both situations are possible.

In this document we do not discuss the efficient solution of (3.3). However, even the best case of orthogonal $\mathcal{T}$, the error in the dirty image is of the same magnitude as the error in the coefficients, which are formed as weighted visibilities. Thus, the error is not only determined by the uncertainty in the measured visibilities, but also but also by the error in the computed weight factors. In order to control it, we need to control the error in the weights.

Let $s_\circ$ be the solution to (3.3) and $\kappa_2(M_\Omega)$ be the condition number of $M_\Omega$ (both for the $\|\cdot\|_2$-norm). If $\hat{s}$ denotes the computed weights and, as we assume that only results with "small" residual norm are accepted, we *conjecture* that

$$\frac{\|s_\circ - \hat{s}\|_2}{\|s_\circ\|_2} \lesssim \kappa_2(M_\Omega) \left( \frac{\|\Delta\psi_\Omega\|_2}{\|\psi_\Omega\|_2} + \frac{\|\Delta M_\Omega\|_2}{\|M_\Omega\|_2} \right) \lesssim \kappa_2(M_\Omega) \delta_\Omega. \quad (3.4)$$

In this case, the error is dominated by the uncertainty in $\Psi_\Omega$, which, when of the order of $\delta_\Omega$, contains a solution with zero residual. The details need to be worked out, but Eq. (3.4) suggests that the rows of $M_\Omega$ (the measurements) need to be selected not only such that their span is close to $\psi_\Omega^*$, but also in such way that the condition number $\kappa_2(M_\Omega)$ does not become too large.[7]

# 4 Numerical examples

## 4.1 The setting

In this section, we consider the simplified case described on Section 3. In order to specify the visibilities analytically, we define intensity distributions $I(\ell)$ that have an analytic Fourier transform. In particular, we consider the sum of Gaussians:

$$I(\ell) = \sum_k a_k e^{-c_k(\ell - p_k)^2}, \quad \text{with} \quad c_k = \frac{1}{2\sigma_k^2}, \quad (4.1)$$

and

$$\mathcal{F}\{I(\ell)\} = \sum_k a_k \sqrt{\frac{\pi}{c_k}} e^{-\frac{\pi^2 u^2}{c_k}} e^{2\pi i u p_k}. \quad (4.2)$$

---

[7]Furthermore, the measurement vectors might be selected by the quality of the measured visibility, i.e. its variance.

For a smooth distribution $I_1(\ell)$, we choose $\{a_1, a_2\} = \{60, 100\}$, $\{\sigma_1, \sigma_2\} = \{0.04, 0.02\}$, and $\{p_1, p_2\} = \{-0.4, 0.2\}$. For a more point source like situation $I_2(\ell)$, we choose $\{a_1, a_2, a_3\} = \{70, 100, 60\}$, $\{\sigma_1, \sigma_2, \sigma_3\} = \{0.005, 0.01, 0.005\}$, and $\{p_1, p_2, p_3\} = \{-0.4, -0.05, 0.3\}$.

We quantify the error in the *computed* brightness distribution $I_{Method}(\vec{\ell})$ by

$$E_{Method} = \frac{\|I(\vec{\ell}) - I_{Method}(\vec{\ell})\|_2}{\|I(\vec{\ell})\|_2} \,. \tag{4.3}$$

However, a smaller error does not necessarily mean a better result. The dirty image is the true intensity convolved with a PSF. A clean image that should recover the true intensity is obtained by subsequent deconvolution.

## 4.2 The principle and low-rank approximations

In this section, we show how a *low-rank approximation* to the image is constructed in an arbitrary basis. We choose a basis that is *convenient to demonstrate the idea*, but not suitable for general use. In particular, we use the Hadamard-Walsh basis [3], as (1) a fast transformation requiring only $\mathcal{O}(P \cdot \log_2 P)$ operations exists, and (2) it is orthogonal.[8]

We define a basis for our approximation subspace $\mathcal{S} = \text{span}\{\psi_\Lambda\}$, by choosing a subset of the Hadamard-Walsh basis $\{\psi_\Omega\}$, $\Lambda \subseteq \Omega$. We discretize the sky with $P = 1{,}024$ points and choose a normalized version of every 16th row of the Hadamard matrix $H_P$. This is equivalent to using the rows of $H_P$ as our basis $\{\psi_\Omega\}$ for the entire image space and setting all coefficients for $\Omega \notin \Lambda = \{0, 16, 32, \ldots\}$ to zero.[9]

In a more practical setting, we would construct the basis of $\mathcal{S}$ based on the measurement vectors, which are given for specific antenna positions and responses, such that (3.3) has a small residual. However, to illustrate the idea, we further simplify the situation depicted in Section 3 by assuming $A_b(\vec{\ell}) = 1$.[10] In this case, for the $\|\cdot\|_2$-norm, the weight factors $\{\sigma_b\}$ are determined by an adjoint NDFT. By assuming the $u$-values are on the grid, which is defined by the discretization of $\ell$, without loss of generality, we can use an FFT to compute the weight factors. Furthermore, for each basis vector, we assume that we measured the visibilities for the $u$-values that correspond to the $\eta \in \mathbb{N}$ largest coefficients of the discrete Fourier spectrum. In fact, we only use these $\eta$ coefficients, which corresponds to restricting the sum in (3.3) to $2\eta$ terms. We remark that *all those assumptions and simplifications are made to demonstrate the idea for the specific choice of basis* and do not have to be made in general.

In this simplified setting, each $\psi_\Lambda$ is approximated by a finite Fourier sum as depicted in Fig. 4.1 for $\psi_{128}$.[11] As the residual norm, $\|\psi_{128} - \hat{\psi}_{128}\|$, is 0.165 and 0.092 for $\eta = 5$ and $\eta = 15$, respectively, the basis vector is actually *not* very well approximated. Despite the relatively bad approximation, especially for some of the other basis elements, we obtain reasonable results for the Hadamard-Walsh coefficients and the reconstructed images, which are shown in Fig. 4.2.

The overall compute cost are as follows: the computation of the Hadamard-Walsh coefficients requires $\dim(\mathcal{S}) \cdot 10\eta = \frac{10}{16}\eta P$ real floating point operations (flops). Thus, the image reconstruction, including the inverse transform, requires only $\mathcal{O}(P \cdot \log_2 P)$ flops.

---

[8] That is, $\psi_\Omega = \psi_\Omega^* = \psi_\Omega^D$.

[9] We used `scipy.linalg.hadamard(P)` to construct $H_P$ and thus start row indexing with zero.

[10] As we discuss below, as each coefficient is a linear combination of only a few visibilities, we are not able to capture finer structures very well.

[11] In fact, we convolved $\psi_{128}$ with $(\frac{1}{4}, \frac{1}{4}, \frac{1}{4}, \frac{1}{4})$ before finding the Fourier coefficients.

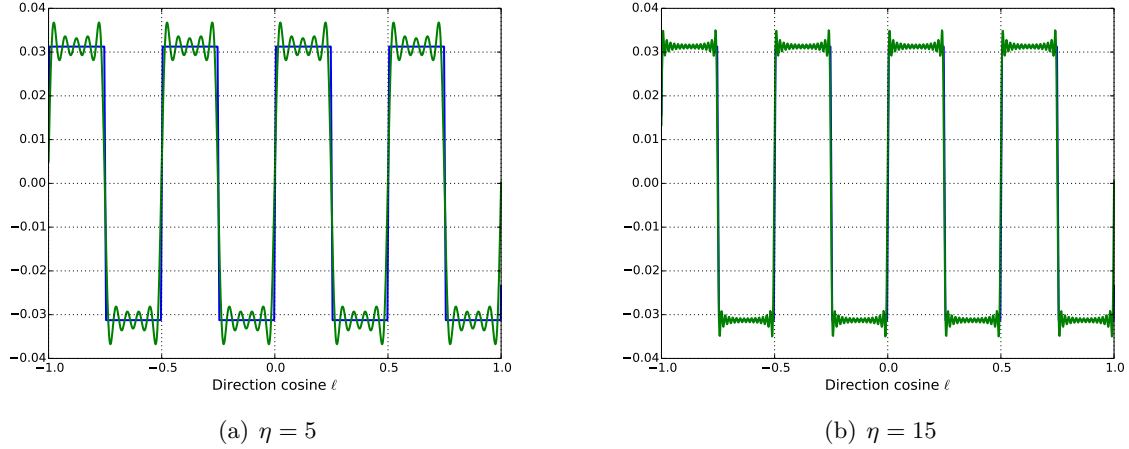

Figure 4.1: Approximation of $\psi_{128}$ by $\hat{\psi}_{128}$. In this case, $\hat{\psi}_{128}$ is a sum of $\eta$ Fourier components.

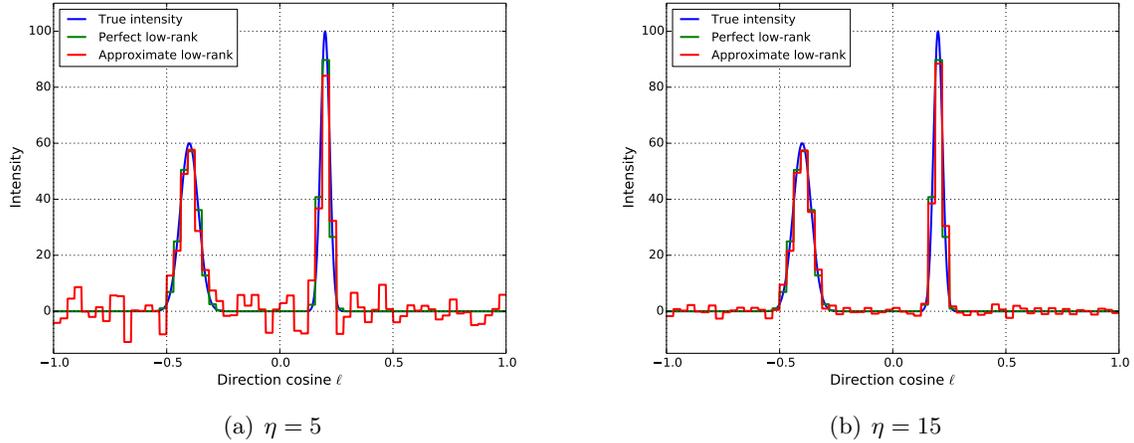

Figure 4.2: The true intensity is given by $I_1(\ell)$. For $\eta = 5$ and $\eta = 15$, the error $E_{HW}$ of 0.36 and 0.25, respectively, and must be compared to the approximation error $E_{\mathcal{S}} = 0.24$ for the perfect image $I_{\mathcal{S}}(\vec{\ell})$ in $\mathcal{S}$. For a better choice of a basis and $\mathcal{S}$, the images would look visually more appealing.

To summarize the low-rank approximation procedure: for a given basis $\{\psi_\Omega\}$, we find the weights in (3.3) and corresponding residuals. All the vectors that are well-approximated (residuals smaller than a certain tolerance), define our approximation subspace $\mathcal{S}$. The coefficients are found by forming linear combinations of the measured visibilities, before a final inverse transform gives the image approximation. The choice of basis is crucial, as $\dim(\mathcal{S}) \ll P$ might be problematic if $\mathcal{S}$ does not contain a good approximation to the true image (i.e., the coefficients that are not obtained are not close to zero). If we have further knowledge about the coefficients, we are able to reduce the computations even further by only considering a subspace $\mathcal{V} \subset \mathcal{S}$.

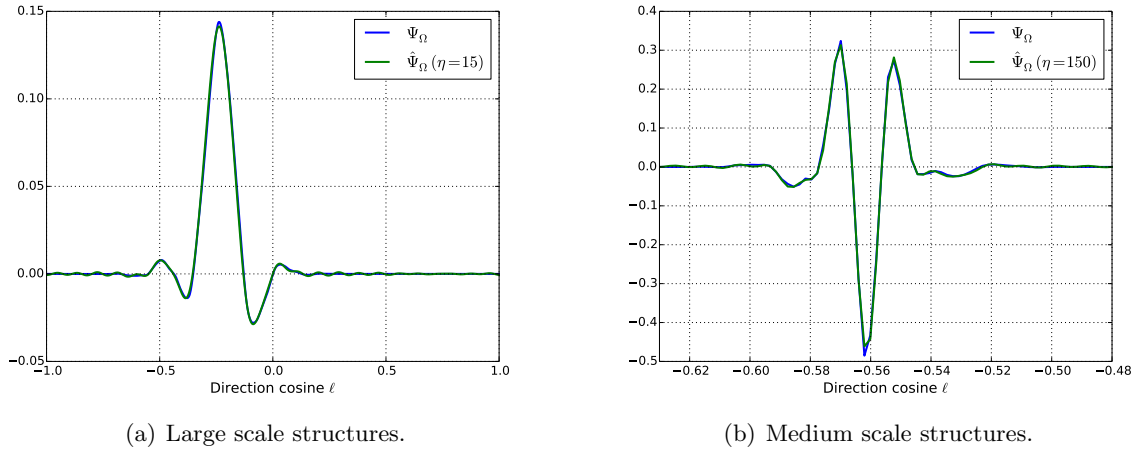

(a) Large scale structures.  (b) Medium scale structures.

Figure 4.3: Naturally, antennas that are sensitive in all directions, $A_b(\vec{\ell}) = 1$, are good to provide information of large-scale structures of the sky, while more localized information is obtained from antennas that are mostly sensitive to the direction of interest. (a) If $\|\psi_\Omega - \hat{\psi}_\Omega\|/\|\psi_\Omega\| = 0.035$ is considered sufficiently accurate, only 15 visibilities are combined to obtain the corresponding wavelet coefficient. (b) For finer structures, to obtain a similar approximation accuracy ($\|\psi_\Omega - \hat{\psi}_\Omega\|/\|\psi_\Omega\| = 0.048$), already 10 times as many visibilities must be considered. For even finer details, this becomes inefficient.

## 4.3  Wavelet transforms

Wavelets promise to provide suitable bases.[12] They have many desirable features: (1) *fast transformations* requiring only $\mathcal{O}(P)$ time exists; (2) they can be *tailored to the specific domain*; (3) they can be "smooth" and are localized; (4) they have an *underlying continuous transform* and come with a vast amount of theory; and, lastly, (5) since a basis is constructed by a scaling and translating a mother wavelet, it is possible to reduce the amount of computations needed to determine the weight factors (e.g., across different frequency bands and for translations). As *wavelets have clear advantages over Fourier basis to represent sharp peaks*, radio astronomical images in a suitable wavelet domain should be relatively sparse. All those facts, make them an interesting family of bases to study.

### 4.3.1  Why the $A$-terms (and $w$-terms) matter

In the previous section, we assumed $A_b(\vec{\ell}) = 1$, i.e. antennas that observe the entire sky from all directions equally well. However, such (idealized) antennas are problematic, when the chosen basis $\{\psi_\Omega\}$ consists of vectors with rather compact support. The problem is best illustrated when assuming $w = 0$. For $A_b(\vec{\ell}) = 1$ and a specific $\psi_\Omega$, weight factors correspond to Fourier components of $\psi_\Omega$. However, if $\psi_\Omega$ has very compact support, many Fourier components must be considered to get a good approximation. Consequently, many visibilities must be combined to get an approximation to the wavelet coefficient. This would be inefficient. The phenomenon is illustrated in Fig. 4.3 for two basis vectors of the `sym6` symlets and $P = 1{,}024$. It becomes inefficient to obtain the coefficients for wavelets that capture the more detailed structures, Fig. 3(b), as many visibilities need to be combined.[13] However, for large scale structures, Fig. 3(a), only a few visibilities must be combined to compute the wavelet coefficient, which is very efficient.

---

[12]Throughout this section, we used the *PyWavelets* python package [7].

If we have the antenna beams $A_b(\vec{\ell})$ that cannot be approximated as being equal to one, or non-zero $w$-terms, we need to find a way to approximate basis elements $\psi_\Omega$ by a finite sum of $A_b(\vec{\ell}) \circ e^{-2\pi i(u_b\vec{\ell}+w_b\vec{n})}$ with only a few terms. We want to provide some intuition, why we believe this can be possible, and why certain wavelets might provide a good basis.

Both the $A$-term, for now assumed to have compact support[14], and the $w$-terms correspond to a "smearing" in the Fourier domain. That is, *we do not measure a single Fourier component anymore*, but all Fourier components in a frequency interval with a certain weighting. (The $AW$-projections reflect this by applying convolutions to the visibilities.) But such compactly supported Fourier spectra actually constitute wavelets. As a simple example, if $A_b$ has Gaussian shape, $e^{-a(\ell-\ell_o)^2}$, $a \in \mathbb{R}^+$, $\ell_o \in [-1,1]$, and $w = 0$, a *single* antenna pair measures a *Gabor wavelet* coefficient. However, Gabor wavelets are non-orthogonal and the question of completeness of the representation must be addressed [4].[15] Furthermore, we have the problem of having our data points not on a "uniform grid" and some form of gridding would be necessary.

Other wavelets are *approximated* by suitable design of antenna beams, $A_b(\vec{\ell})$.[16] Similarly, when $A_b(\vec{\ell})$ and baselines are given, we might *construct* a suitable wavelet such that it can be represented by a sum of a few measurement vectors. In any case, the sum in (3.3) is severely restricted by leaving out baselines for which $A_b(\vec{\ell})$ is close to zero in the region of the wavelet support; that is, we only take antennas into account that are sensitive to the region of the sky that is important to determine the wavelet coefficient.

Purely based on this intuition, we believe that specific wavelets that do not capture details below the capabilities of the instrument, can be represented by a limited sum of measurement vectors; thereby allowing the efficient computation of the corresponding wavelet coefficient. We are currently working on a more detailed analysis.

### 4.3.2 Reconstructing images with wavelets transforms

Although we discussed the important role of the $A$-terms (and $w$-terms), we once again assume $A_b(\vec{\ell}) = 1$ and $w = 0$. This is solely done, as in this case the *visibilities are given analytically*. In contrast, for a more realistic scenario, the visibilities would have to be determined by a simulation. As described above, our assumptions imply that the wavelet coefficients for the detailed structures are linear combinations of many visibilities. This is inefficient, but in this section, we only want to show that it works. Furthermore, we use a rather dense sampling of the $u$-space, by setting $Y = 35$ antennas at positions $\frac{1}{8}k^2$, $k = 0, 1, \ldots, Y - 1$ (using normalized wavelength $\lambda = 1$). The dense sampling implies that all elements of basis $\{\psi_\Omega\}$ are well approximated.[17] (We used tolerance $\delta_\Omega = 10^{-5}$.)

---

[13] However, it is efficient to compute the weight factors, as all the translates in a wavelet basis are simple phase shifts for the weight factors.

[14] Else, it might be considered a weighted sum $A_b(\ell) = \sum_j \beta_j A_j(\ell - \ell_j)$ of normalized antenna responses $A_j$ with compact support.

[15] We are not aware of studies applying windowed Fourier transforms to radio astronomical imaging.

[16] This can be done via beamforming.

[17] The sampling in $u$-space, and a general scenario the $(u, v, w)$-coverage, is absolutely crucial to the success of the approach, as it is for the classical Fourier-based reconstruction.

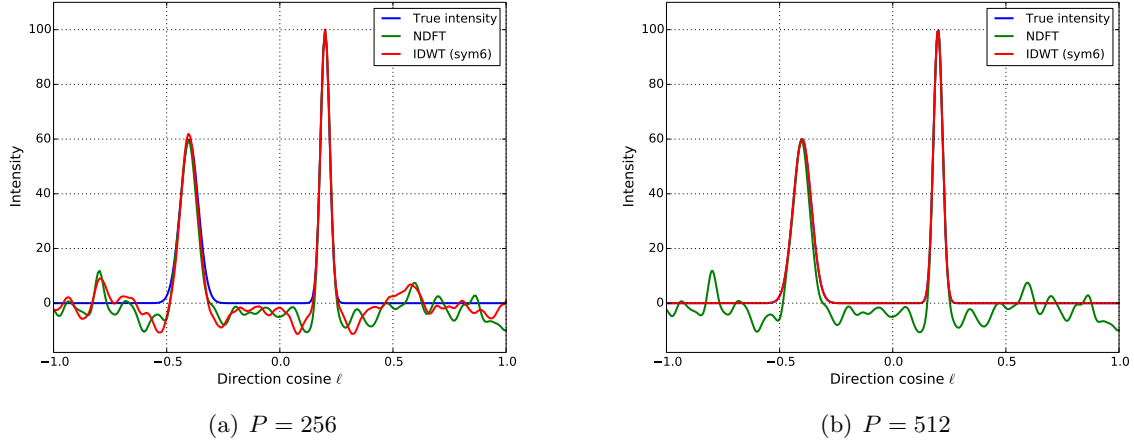

(a) $P = 256$

(b) $P = 512$

Figure 4.4: The true intensity is given by $I_1(\vec{\ell})$. The results are normalized such that $\|I_1(\vec{\ell})\|_1 = \|I_{NDFT}(\vec{\ell})\|_1 = \|I_{IDWT}(\vec{\ell})\|_1$. In this case, $E_{NDFT} = 0.285$, $E_{IDWT} = 0.263$ ($P = 256$), and $E_{IDWT} = 1.2 \cdot 10^{-12}$ ($P = 512$).

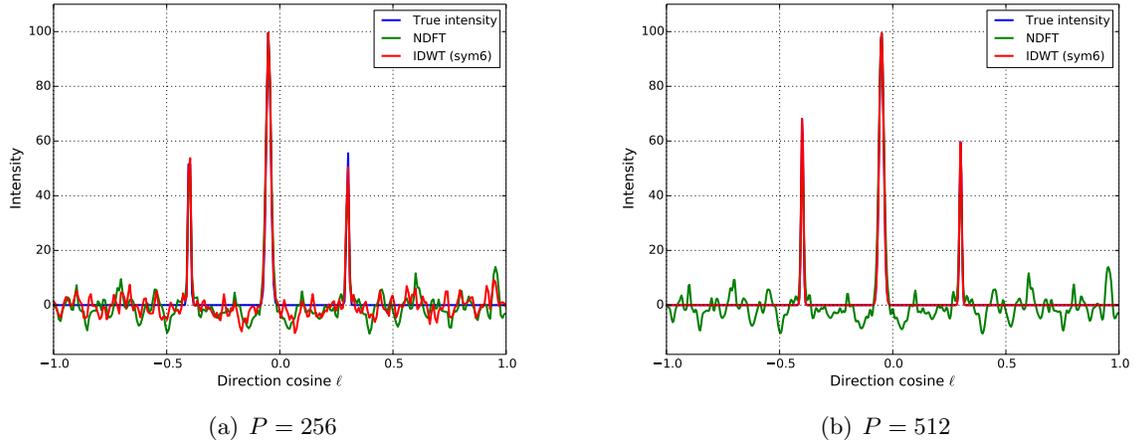

(a) $P = 256$

(b) $P = 512$

Figure 4.5: The true intensity is given by $I_2(\vec{\ell})$. The results are normalized such that $\|I_2(\vec{\ell})\|_1 = \|I_{NDFT}(\vec{\ell})\|_1 = \|I_{IDWT}(\vec{\ell})\|_1$. In this case, $E_{NDFT} = 0.378$, $E_{IDWT} = 0.293$ ($P = 256$), and $E_{IDWT} = 1.2 \cdot 10^{-3}$ ($P = 512$).

We also mentioned the importance to restrict the sum in (3.3) and that, with the above assumptions, weights for a translated basis element are determined by a simple phase shift. We did not make any use of this and solved (3.3) in a brute-force way instead. To keep the computational cost within limits, we restricted $P$ to 256 and 512.

For the two intensity distributions $I_1(\vec{\ell})$ and $I_2(\vec{\ell})$, the results are presented in Figs. 4.4 and 4.5, respectively. With a single wavelet transform, we are able to reconstruct the intensity quite well. *However, in this section with its simplified assumptions, we only demonstrate a proof of concept: images are created by an inverse discrete wavelet transform!* The details, optimization,

Table 5.1: Traditional vs. generalized approach.

| | Traditional | Generalized |
|---|---|---|
| Basis | $e^{-2\pi i(u_{ij}\ell + v_{ij}m)}$ | $\sum_{i,j,k} \sigma_{i,j,k} A_{ij,k}(\ell,m)\, e^{-2\pi i(u_{ij}\ell + v_{ij}m + w_{ij}n)} \approx \Psi_\Omega^D$ |
| Dirty image | Inverse Fourier transform of sampled visibilities $V(u,v)$: $\mathcal{F}^{-1}(V(u,v))$ Performed by gridding + FFT. | Any inverse transform, depending on basis $\{\Psi_\Omega\}$: $\mathcal{T}^{-1}(C(\Psi_\Omega))$ (e.g., FWT) |
| $AW$-terms | Corrections via convolutions | Included in the construction of the basis |
| Clean image | Deconvolution with PSF | Deconvolution with PSF defined by weights |

and practical issues must still be investigated.

## 5 Open questions and future research

Before concluding, we give a short comparison of the traditional and the generalized approach in Table 5.1. *The generalized procedure is far from being fully developed.* Before it can be used in practice, a number of questions need to be addressed. At this point, we collect a few of them:

- For a given antenna configuration and the radio astronomical application domain, which basis is "optimal"? Or, can we co-design the placements and beams of antennas and the image basis? Is there a design that allows for sparse weighting vectors?
- How to determine weights for a specific choice of basis and antenna configuration *efficiently*?
- For which use cases could the approach be beneficial? (E.g., low-resolution all-sky imaging.)
- How do errors (discretization, approximation, algorithmic) influence the final result?

In the future, we plan to address these (and more) questions in order to work out the approach in detail for a realistic scenario. In Appendix A, we describe the procedure in matrix form, which is more suitable for a detailed analysis.

## 6 Conclusions

We *generalize* the procedure of obtaining a radio astronomical sky image by an non-uniform discrete Fourier transformation. As a result, in our generalized approach, many transformations are possible, giving raise to *whole new space to explore*. In particular, we demonstrate how an image is obtained by a fast discrete wavelet transform. As such transformations can be tailored to the application domain and performed in linear time, they promise to be one of the ingredients of real-time, wide-field imaging algorithms.

# Acknowledgments

The research leading to these results has been conducted in the frame of the DEEP-ER (Dynamically Exascale Entry Platform - Extended Reach) project, which receives funding from the European Communitys Seventh Framework Programme (FP7/2007-2013) under Grant Agreement n° 610476. The author would like to thank Bram Veenboer (especially for the help with TikZ), John Romein, Yan Grange, Albert-Jan Boonstra, Ronald Nijboer, and Jan Noordam for discussions. Furthermore, I am thankful towards all the people participating in the reading club at ASTRON, as they tried to teach me how a radio telescope works during the last months (if I did not get it right, they are not responsible). Last but not least, I want to thank Tammo Jan Dijkema for reading an earlier version of this manuscript and providing many useful suggestions.

## A   Matrix notation

In this section, we investigate the one-dimensional case, which was described in Section 3, in a different notation. Let $\vec{\imath} = I(\vec{\ell}) \in \mathbb{R}^P$ be the intensity (or image), $\vec{v} = V(\{u_b\}) \in \mathbb{C}^{2B}$ the set of measured visibilities. Eq. (3.1) then becomes

$$\vec{v} = M\vec{\imath}, \tag{A.1}$$

where $M \in \mathbb{C}^{2B \times P}$ is the measurement matrix, with $A_b(\ell_p)e^{-2\pi i u_b \ell_p}$ and $p = 1, 2, \ldots, P$ as its rows. With the pseudoinverse $M^\dagger$, we have a solution $\vec{\imath} = M^\dagger \vec{v}$ (or select any of the solutions $M^\dagger \vec{v} - (I - M^\dagger M)q$ for all $q \in \mathbb{C}^P$). The only problem is that solving (A.1) is generally expensive.

Ideally we would have the following situation: let $T^* \in \mathbb{C}^{P \times P}$ be *any* unitary matrix, often real and orthogonal, and the system

$$\vec{g} = T^*\vec{\imath}, \tag{A.2}$$

solvable in $\mathcal{O}(P)$ or $\mathcal{O}(P \log_2 P)$ time. If we could estimate $\vec{g} \in \mathbb{C}^P$ in comparable time, the imaging problem could be solved very efficiently. Similarly, if

$$\vec{c} = \hat{T}^*\vec{\imath} \tag{A.3}$$

for known $\vec{c} \in \mathbb{C}^P$ and $\hat{T}^* \approx T^*$, we can solve $\vec{c} = T^*\hat{\imath}$ to obtain an approximation to the image.

Let $\vec{w}_j \in \mathbb{C}^{2B}$, $j = 1, 2, \ldots, P$, be a vector of unknown weights. By forming

$$\vec{w}_j^*\vec{v} = \vec{w}_j^*M\vec{\imath}, \tag{A.4}$$

and comparing to (A.3), we find that if $\vec{w}_j^*M$ approximates well the $j$-th row of $T^*$, $\psi_j^*$, $c_j = \vec{w}_j^*\vec{v}$ is a good approximation to $g_j = \psi_j^*\vec{\imath}$. This statement is quantified by the Cauchy-Schwartz inequality, as

$$|g_j - c_j| = |(\psi_j^* - \vec{w}_j^*M)\vec{\imath}| \leq \|\psi_j^* - \vec{w}_j^*M\|_2 \|\vec{\imath}\|_2. \tag{A.5}$$

If we require a certain accuracy in the coefficient approximation, (A.5) gives an idea on how big $\|\psi_j^* - \vec{w}_j^*M\|_2$ is tolerable.[18] In general, we require

$$\|\psi_j^* - \vec{w}_j^*M\| \leq \delta_j \tag{A.6}$$

to consider $c_j$ and accurate approximation to $g_j$. If (A.6) does not holds, we set $c_j$ to zero.

---

[18] By (A.5), the approximation error can be large in parts of the shy where there are no strong sources.

Without any loss of generality, assume (A.6) holds for $j = 1, 2, \ldots, S$ and not for $j = S+1, \ldots, P$. Let $W^* \in \mathbb{C}^{S \times 2B}$ have rows $w_j^*$ and partition (A.3) as follows

$$\begin{pmatrix} \vec{c}_1 \\ 0 \end{pmatrix} = \begin{pmatrix} \hat{T}_1^* \\ \hat{T}_2^* \end{pmatrix} \vec{\imath} = \underbrace{\begin{pmatrix} W^*M \\ T_2^* \end{pmatrix}}_{\hat{T}} \vec{\imath} \tag{A.7}$$

where $\vec{c}_1 \in \mathbb{C}^S$, $\hat{T}_1^* \in \mathbb{C}^{S \times P}$, $\hat{T}_2^* \in \mathbb{C}^{(P-S) \times P}$. In (A.7), $T_2^* \in \mathbb{C}^{(P-S) \times P}$ is the part of $T^*$ that is formed by all the $\psi_j^*$ that do not full-fill (A.6).

Our approximate image $\hat{\imath}$ in $\mathcal{S} = \text{span}\{\psi_j : j = 1, 2, \ldots, S\}$ is obtained by solving

$$\begin{pmatrix} \vec{c}_1 \\ 0 \end{pmatrix} = \begin{pmatrix} T_1^* \\ T_2^* \end{pmatrix} \hat{\imath}, \tag{A.8}$$

in $\mathcal{O}(P)$ or $\mathcal{O}(P \log_2 P)$ time, i.e. $\hat{\imath} = T_1 \vec{c}_1$. We introduce an error by replacing $\hat{T}_1$ by $T_1$. However,

$$\frac{\|(T_1 - \hat{T}_1)\vec{c}_1\|_2}{\|\hat{\imath}\|_2} \leq \frac{\|T_1 - \hat{T}_1\|_2 \|\vec{c}_1\|_2}{\|T_1 \vec{c}_1\|_2} \leq \|T_1^* - W^*M\|_2. \tag{A.9}$$

But, $\|T_1^* - W^*M\|_2$ is easily bounded as (A.6) holds for every row.

There are a number of questions that need to be addressed: (1) The existence of $W$ with $S \approx P$; (2) the sparsity of $W$ such that forming $\vec{c}_1 = W^* \vec{v}$ is performed in $\mathcal{O}(P)$ or $\mathcal{O}(P \cdot \log_2 P)$ operations. Both questions are highly coupled with the specific form of measurement matrix $M$ and transformation $T$. In this paper, we do not address these questions. We simply state the problem again: *Does there exists a unitary $T \in \mathbb{C}^{P \times P}$ and a sparse matrix $W^* \in \mathbb{C}^{S \times 2B}$, where $\vec{g} = T^* \vec{\imath}$ is solvable in $\mathcal{O}(P \cdot \log_2 P)$ time, and for any subset of $S$ rows of $T^*$ (ideally, $S \approx P$), $T_1^* \in \mathbb{C}^{S \times P}$, not only $\|T_1^* - W^*M\| \leq \delta$, for some tolerance $\delta$, but also $W^* \vec{c}_1$ is performed in $\mathcal{O}(P \cdot \log_2 P)$ time?*

Also, why can the sketched approach be fast, if the determination of $W$ is potentially expensive? The answer is that $W$ is determined *independent of the image* that is created. That is, the weights are determined before an observation starts ("offline"). The idea is as follows: before an observation, we know how $M$ will involve during time. More rows are added as measurements are performed. For each moment of the observation, we determine a matrix $W$ such that $\vec{c}_1 = W^* \vec{v}$, where values of $\vec{v}$ and rows $W$ are added for the measurements. If we have domain-specific knowledge that some coefficients of $\vec{c}_1$ are close to zero these coefficients are not computed, also the weights $W$ (the associated columns) do not need to be determined. In this way, when starting an observing, we obtain a better and better image as time evolves. In a steady state, depending on how much measurement from the past are included in the image creation (old measurements are removed from $M$ and $\vec{v}$), the observation is showing the sky integrated during a certain time window. *During the observation* ("online"), as $W$ is known and assumed to be sparse, *the image creation is computationally efficient.*